\documentclass[aps,twocolumn,pra,superscriptaddress,showpacs,tightenlines]{revtex4-1}
\usepackage{amsmath}
\usepackage{graphicx}
\usepackage{color}
\usepackage{amsfonts}

\begin{document}
\title{Entangling two macroscopic mechanical mirrors in a two-cavity optomechanical system}
\author{Jie-Qiao Liao}
\affiliation{CEMS, RIKEN, Saitama 351-0198, Japan}
\author{Qin-Qin Wu}
\affiliation{Department of Physics and Electronics, Hunan
Institute of Science and Technology, Yueyang 414000, China}
\author{Franco Nori}
\affiliation{CEMS, RIKEN, Saitama 351-0198, Japan}
\affiliation{Department of Physics, The University of Michigan, Ann Arbor, Michigan 48109-1040, USA}
\date{\today}
\begin{abstract}
We propose a simple method to generate quantum entanglement between two macroscopic mechanical resonators in a two-cavity optomechanical system.
This entanglement is induced by the radiation pressure of a single photon hopping between the two cavities.
Our results are analytical, so that the entangled states are explicitly shown. Up to local operations,
these states are two-mode three-component states, and hence the degree of entanglement can be well quantified by the concurrence.
By analyzing the system parameters, we find that, to achieve a maximum average entanglement, the system should work
in the single-photon strong-coupling regime and the deep-resolved-sideband regime.
\end{abstract}

\maketitle

Quantum entanglement~\cite{Schrodinger1935,Horodecki2009}, as a cornerstone of quantum physics, plays an important role in the foundation of quantum theory
and also has potential applications in quantum technology, such as quantum information science~\cite{Jones2012} and quantum metrology~\cite{Giovannetti2004}.
In particular, how to prepare macroscopic mechanical entanglement is of high interest and significance,
because such macroscopic entanglement might provide explicit evidence for quantum phenomena~\cite{Schwab2005}
and even might possibly help us to clarify the quantum-to-classical transition,
as well as the boundary between classical and quantum worlds~\cite{Zurek1991}. Recently, much attention has been paid to the
creation of quantum entanglement in macroscopic mechanical systems. Some proposals have been brought forward to generate quantum
entanglement in various mechanical resonators~\cite{Mancini2002,Eisert2004,Pinard2005,Xue2007,Paz2008,Hartmann2008,Vacanti2008,Jost2009,Huang2009,Ludwig2010,Borkje2011,Zhou2011,Joshi2012,Xu2013,Walter2013,Tan2013}.

In general, we can classify these state-preparation proposals into two categories, according to the coupling channels:
either direct coupling or indirect coupling. In the former case, the two mechanical resonators are coupled to each other directly.
In the latter case, some kind of an intermediate is needed to induce an effective interaction
between the two mechanical resonators. Therefore, the intermediate link should be able to couple with the mechanical resonators.
In this sense, cavity optomechanical systems~\cite{Kippenberg2008,Marquardt2009,Aspelmeyer2013} can provide a natural platform to induce an interaction between mechanical resonators
because there is an intrinsic coupling mechanism between optical and mechanical degrees of freedom.
Motivated by this feature, in this paper we propose to study the generation of macroscopic mechanical
entanglement in a two-cavity optomechanical system. This system is composed of two coupled optomechanical cavities.
In each cavity, the electromagnetic fields couple to the mechanical motion
of one moving end mirror. The connection between the two cavity fields is built through a photon-hopping interaction. This
photon connection will induce an entanglement between the two mechanical mirrors. We note that some previous studies have considered various entanglements
in optomechanical systems~\cite{Ferreira2006,Vitali2007,Paternostro2007,Genes2008,Wang2013,Tian2013,He2013}.

In particular, we will focus on the single-photon strong-coupling regime~\cite{Rabl2011,Nunnenkamp2011,Chenp2011,Liao2012,He2012,Liao2013,Liu2013,Kronwald2013,XuandLaw2013},
in which the radiation pressure of a single photon can produce observable effects.
In this regime, people have found that strong photon nonlinearity at the few-photon level (e.g., photon blockade)
can be induced by the radiation-pressure coupling~\cite{Rabl2011,Liao2013,Liu2013,Kronwald2013};
moreover, resolved phonon sidebands and frequency shifts can be observed in the photon emission and scattering spectra~\cite{Liao2012}.
So, a natural question is whether a single photon can also induce a considerable
entanglement between the two mechanical resonators in the single-photon strong-coupling regime.
Below, we will address this question by analytically solving the dynamics of the system.

\begin{figure}[tbp]
\center
\includegraphics[bb=44 673 371 742, width=3.3 in]{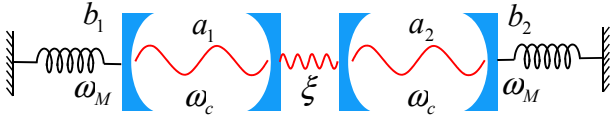}
\caption{(Color online) Schematic diagram of the two-cavity optomechanical system.
Each Fabry-Perot cavity is formed by a fixed end mirror and a moving one. The electromagnetic
fields in the cavity couple to the mechanical motion of the moving mirror through the radiation-pressure interaction, and
the fields in the two cavities couple to each other via a photon-hopping interaction.}
\label{setup}
\end{figure}

Specifically, we consider a two-cavity optomechanical system, which
consists of two optomechanical cavities (Fig.~\ref{setup}).
Each cavity is formed by a fixed end mirror and a moving one.
We focus on a single-mode electromagnetic field in each cavity. This field couples to the mechanical motion of
the moving mirror via the radiation pressure coupling. In addition,
the fields in the two cavities couple to each other via a photon-hopping
interaction. Without loss of generality, we assume that the two
optomechanical cavities are identical. The Hamiltonian of the system
is ($\hbar=1$)
\begin{eqnarray}
H_{S}&=&\sum_{j=1,2}\left[\omega_{c}a_{j}^{\dagger}a_{j}+\omega_{M}b_{j}^{\dagger}b_{j}
-g_{0}a_{j}^{\dagger}a_{j}(b_{j}^{\dagger}+b_{j})\right]\notag\\
&&-\xi(a_{1}^{\dagger}a_{2}+a_{2}^{\dagger}a_{1}),\label{Hamiltonianorig}
\end{eqnarray}
where $a_{j}$ $(a_{j}^{\dagger})$ and $b_{j}$ $(b_{j}^{\dagger})$ are,
respectively, the annihilation (creation) operators of the single-mode
cavity field and the mechanical motion of the moving mirror in the $j$th ($j=1,2$) optomechanical cavity, with
respective resonant frequencies $\omega_{c}$ and $\omega_{M}$. The
parameter $g_{0}=\omega_{c}x_{\text{zpf}}/L$ is the single-photon optomechanical coupling
strength, where $x_{\text{zpf}}=\sqrt{1/(2M\omega_{M})}$ is the zero-point fluctuation of the moving mirror with
mass $M$, and $L$ is the rest length of the cavity. The parameter $\xi$ is
the photon-hopping coupling strength between the two cavities. We note that some previous studies have considered multicavity optomechanical systems with only one mechanical resonator~\cite{Miao2009,Zhao2009,Dobrindt2010,Ludwig2012,Stannigel2012,Komar2013,Lu2013}.

To solve the Hamiltonian $H_{S}$, we first introduce the transformation
$V_{1}=\exp[\frac{\pi}{4}(b_{1}^{\dagger}b_{2}-b_{2}^{\dagger}b_{1})]$,
and the transformed Hamiltonian then becomes
\begin{eqnarray}
H_{1}&=&V_{1}H_{S}V_{1}^{\dagger}\notag\\
&=&\omega_{c}(a_{1}^{\dagger}a_{1}+a_{2}^{\dagger}a_{2})+\omega_{M}(b_{1}^{\dagger}b_{1}+b_{2}^{\dagger}b_{2})\nonumber\\
&&+g(a_{1}^{\dagger}a_{1}+a_{2}^{\dagger}a_{2})(b_{1}^{\dagger}+b_{1})+g(a_{2}^{\dagger}a_{2}-a_{1}^{\dagger}a_{1})(b_{2}^{\dagger}+b_{2})\notag\\
&&-\xi(a_{1}^{\dagger}a_{2}+a_{2}^{\dagger}a_{1}),\label{H1orig}
\end{eqnarray}
where $g\equiv-g_{0}/\sqrt{2}$. In the absence of cavity photon decay, the total cavity photon number in this system
is a conserved quantity because of $[a_{1}^{\dagger }a_{1}+a_{2}^{\dagger }a_{2},H_{1}]=0$, and hence we can restrict the system within a subspace with a definite total photon number. In this work, we will
consider the single-photon case, i.e., $a_{1}^{\dagger}a_{1}+a_{2}^{\dagger}a_{2}=1$, then the Hamiltonian~(\ref{H1orig}) is reduced to
\begin{eqnarray}
\label{e2}
H_{1}&=&\omega_{c}+\omega_{M}b_{1}^{\dagger}b_{1}+g(b_{1}^{\dagger}+b_{1})-\xi(a_{1}^{\dagger}a_{2}+a_{2}^{\dagger}a_{1})\notag\\
&&+\omega_{M}b_{2}^{\dagger}b_{2}+g(a_{2}^{\dagger}a_{2}-a_{1}^{\dagger}a_{1})(b_{2}^{\dagger}+b_{2}).
\end{eqnarray}
Now the mode $b_{1}$ decouples from other modes. Using the transformation
$V_{2}=\exp[\frac{\pi}{4}(a_{1}^{\dagger}a_{2}-a_{2}^{\dagger}a_{1})]$,
the Hamiltonian $H_{1}$ can be further transformed to
\begin{eqnarray}
\label{e3}
H_{2}&=&V_{2}H_{1}V_{2}^{\dagger}=H_{\textrm{I}}+H_{\textrm{II}},
\end{eqnarray}
where $H_{\textrm{I}}$ and $H_{\textrm{II}}$ are defined by
\begin{subequations}
\begin{align}
H_{\textrm{I}}=&\omega_{c}+\omega_{M}b_{1}^{\dagger}b_{1}+g(b_{1}^{\dagger}+b_{1}),\\
H_{\textrm{II}}=&\xi(a_{2}^{\dagger}a_{2}-a_{1}^{\dagger}a_{1})+\omega_{M}b_{2}^{\dagger}b_{2}\nonumber\\
&+g(a_{1}^{\dagger}a_{2}+a_{2}^{\dagger}a_{1})(b_{2}^{\dagger}+b_{2}).
\end{align}
\end{subequations}
These two parts $H_{\textrm{I}}$ and $H_{\textrm{II}}$ are commutative, i.e., $[H_{\textrm{I}},H_{\textrm{II}}]=0$.
In particular, the Hamiltonian $H_{\textrm{I}}$ can be diagonalized with the displacement transformation $V_{3}=\exp[\frac{g}{\omega_{M}}(b_{1}^{\dagger}-b_{1})]$ as follows:
\begin{eqnarray}
\label{HamitH3}
H_{3}=V_{3}H_{\textrm{I}}V_{3}^{\dagger}=\omega_{M}b_{1}^{\dagger}b_{1}+\omega_{c}-\frac{g^{2}}{\omega_{M}}.
\end{eqnarray}
For the Hamiltonian $H_{\textrm{II}}$, if we consider the case $\omega_{M}=2\xi\gg g$, and in the single-photon case, it can be
written as the Jaynes-Cummings Hamiltonian under the rotating-wave
approximation,
\begin{eqnarray}
H_{\textrm{II}}&=&\frac{\omega_{M}}{2}(\vert 0\rangle_{a_{1}}\vert 1\rangle_{a_{2}}\;_{a_{2}}\langle 1|_{a_{1}}\langle0|-\vert 1\rangle_{a_{1}}\vert 0\rangle_{a_{2}}\;_{a_{2}}\langle 0|_{a_{1}}\langle 1|),\notag \\
&&+g(\vert 1\rangle_{a_{1}}\vert 0\rangle_{a_{2}}\;_{a_{2}}\langle 1|_{a_{1}}\langle0| b_{2}^{\dagger}+\vert 0\rangle_{a_{1}}\vert 1\rangle_{a_{2}}\;_{a_{2}}\langle 0|_{a_{1}}\langle 1| b_{2})\nonumber\\
&&+\omega_{M}b_{2}^{\dagger}b_{2}.\label{e6}
\end{eqnarray}

Now, we can solve the dynamics of the system Hamiltonian $H_{S}$ in the single-photon subspace, because the two Hamiltonians $H_{3}$ and $H_{\textrm{II}}$ are solvable.
The unitary evolution operator associated with the Hamiltonian $H_{S}$ in the single-photon subspace can be expressed as
\begin{equation}
U(t)=e^{-iH_{S}t}=V_{1}^{\dagger}V_{2}^{\dagger}V_{3}^{\dagger}e^{-iH_{3}t}V_{3}e^{-iH_{\textrm{II}}t}V_{2}V_{1}.\label{unitaryevolop}
\end{equation}
Based on this unitary evolution operator $U(t)$, we can calculate the state of the system
at any time $t$ once the initial state is given.
To induce an entanglement between the two mechanical resonators, we introduce a single photon as an intermediate.
The initial state of the system is assumed to be $\vert\psi(0)\rangle=\vert 1\rangle_{a_{1}}\vert0\rangle_{a_{2}}\vert0\rangle_{b_{1}}\vert0\rangle_{b_{2}}$.
After some tedious calculations, the state of the system at time $t$ can be obtained as
\begin{eqnarray}
\vert\psi(t)\rangle&=&U(t)\vert\psi(0)\rangle\nonumber\\
&=&\frac{1}{2}e^{-i\Theta(t)}D_{b_{1}}[\beta(t)/\sqrt{2}]D_{b_{2}}[\beta(t)/\sqrt{2}]\nonumber\\
&&\times\left(\vert1\rangle_{a_{1}}\vert 0\rangle_{a_{2}}\left\{\left[\exp(i\omega_{M}t)+\cos(gt)\right]\vert0\rangle_{b_{1}}\vert0\rangle_{b_{2}}\right.\right.\nonumber\\
&&\left.\left.+\frac{i}{\sqrt{2}}\sin(gt)(\vert0\rangle_{b_{1}}\vert 1\rangle
_{b_{2}}-\vert 1\rangle_{b_{1}}\vert0\rangle_{b_{2}})\right\}\right.\nonumber\\
&&\left.+\vert 0\rangle_{a_{1}}\vert 1\rangle_{a_{2}}\left\{\left[\exp(i\omega_{M}t)-\cos(gt)\right]\vert0\rangle_{b_{1}}\vert0\rangle_{b_{2}}\right.\right.\nonumber\\
&&\left.\left.+\frac{i}{\sqrt{2}}\sin(gt)(\vert 0\rangle_{b_{1}}\vert 1\rangle_{b_{2}}-\vert 1\rangle_{b_{1}}\vert0\rangle_{b_{2}})\right\}\right),\label{genstate}
\end{eqnarray}
where the phase factor is
$\Theta(t)=(\omega_{c}+\omega_{M}/2)t+(g^{2}/\omega_{M})[\sin(\omega_{M}t)/\omega_{M}-t]$,
and $D_{b_{j}}[\beta(t)/\sqrt{2}]=\exp[(\beta(t)b_{j}^{\dag}-\beta^{\ast}(t)b_{j})/\sqrt{2}]$ is the
displacement operator with the parameter
$\beta(t)=-(g/\omega_{M})(1-e^{-i\omega_{M}t})$.

Generally, the state $\vert\psi(t)\rangle$ is an entangled state involving two cavity fields and two mechanical modes.
However, we can obtain macroscopic mechanical entanglement by measuring the state of the two cavity fields:

(i) If the cavity-field state is detected in $\vert 1\rangle_{a_{1}}\vert 0\rangle_{a_{2}}$, i.e., the single photon is in the first cavity, then the state of the
two mechanical modes becomes
\begin{eqnarray}
\vert\varphi_{1}(t)\rangle=D_{b_{1}}[\beta(t)/\sqrt{2}]D_{b_{2}}[\beta(t)/\sqrt{2}]\vert\phi_{1}(t)\rangle,\label{e10}
\end{eqnarray}
with
\begin{eqnarray}
\vert\phi_{1}(t)\rangle&=&\mathcal{N}_{1}
\left\{[\exp(i\omega_{M}t)+\cos(gt)]\vert 0\rangle_{b_{1}}\vert 0\rangle_{b_{2}}\right.\nonumber\\
&&\left.+(i/\sqrt{2})\sin(gt)(\vert 0\rangle_{b_{1}}\vert 1\rangle_{b_{2}}-\vert 1\rangle
_{b_{1}}\vert 0\rangle_{b_{2}})\right\},\label{genstatephi1}
\end{eqnarray}
where $\mathcal{N}_{1}=1/\sqrt{2+2\cos(gt)\cos(\omega_{M}t)}$ is the
normalization constant. We can evaluate the probability for detecting this component state $\vert 1\rangle_{a_{1}}\vert 0\rangle_{a_{2}}$  based on Eq.~(\ref{genstate}). This is
\begin{equation}
P_{1}(t)=\frac{1}{2}[1+\cos(gt)\cos(\omega_{M}t)].\label{prob1}
\end{equation}

(ii) If the cavity-field state is detected in $\vert 0\rangle_{a_{1}}\vert 1\rangle_{a_{2}}$, i.e., the single photon is in the second cavity, then the
state of the two mechanical modes is
\begin{eqnarray}
\label{e12}\vert\varphi_{2}(t)\rangle=D_{b_{1}}[\beta(t)/\sqrt{2}]D_{b_{2}}[\beta(t)/\sqrt{2}]\vert\phi_{2}(t)\rangle,
\end{eqnarray}
where
\begin{eqnarray}
\vert\phi_{2}(t)\rangle&=&\mathcal{N}_{2}\left\{[\exp(i\omega_{M}t)-\cos(gt)]\vert 0\rangle_{b_{1}}\vert 0\rangle_{b_{2}}\right.\nonumber\\
&&+\left.(i/\sqrt{2})\sin(gt)(\vert 0\rangle_{b_{1}}\vert 1\rangle_{b_{2}}-\vert 1\rangle_{b_{1}}\vert 0\rangle_{b_{2}})
\right\},\label{genstatephi2}
\end{eqnarray}
with $\mathcal{N}_{2}=1/\sqrt{2-2\cos(gt)\cos(\omega_{M}t)}$.
The probability for this component state is
\begin{eqnarray}
P_{2}(t)=\frac{1}{2}[1-\cos(gt)\cos(\omega_{M}t)].\label{prob2}
\end{eqnarray}

\begin{figure}[tbp]
\center
\includegraphics[bb=12 14 496 536, width=3.3 in]{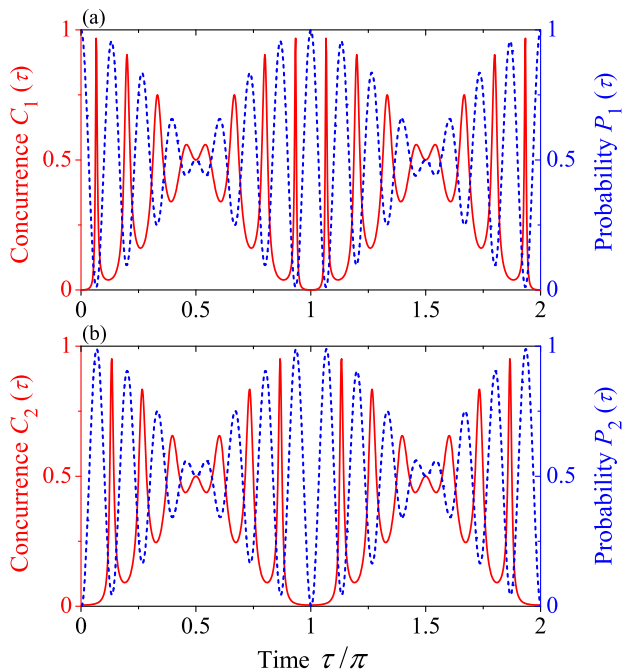}
\caption{(Color online) Dynamics of the concurrence (red solid curves) and the probability (blue short-dashed curves)
for states (a) $\vert\phi_{1}(t)\rangle$ and (b) $\vert\phi_{2}(t)\rangle$.
Here, $\tau\equiv gt$ is the scaled time and $\omega_{M}/g=15$.}
\label{concuandproba}
\end{figure}
So far, we have obtained the quantum entangled states of the two mechanical modes.
Below, we give a proper description and quantification of the entanglement.
Since the displacement operators $D_{b_{1}}[\beta(t)/\sqrt{2}]$ and $D_{b_{2}}[\beta(t)/\sqrt{2}]$
are local operations, the states $|\varphi_{l}(t)\rangle$ and $|\phi_{l}(t)\rangle$ ($l=1,2$) have the same degree of entanglement.
The states $|\phi_{1}(t)\rangle$ and $|\phi_{2}(t)\rangle$ are actually two-mode three-component pure states, so we
can employ the concurrence~\cite{Wootters1998} as the measurement of the degree of entanglement. The respective concurrences for states $|\phi_{1}(t)\rangle$ and $|\phi_{2}(t)\rangle$
are
\begin{subequations}
\label{concurrence}
\begin{align}
C_{1}(t)&=\frac{\sin^{2}(gt)}{2[1+\cos(gt)\cos(\omega_{M}t)]},\\
C_{2}(t)&=\frac{\sin^{2}(gt)}{2[1-\cos(gt)\cos(\omega_{M}t)]}.
\end{align}
\end{subequations}

To evaluate the performance of the entanglement generation, in Fig.~\ref{concuandproba}, we show the evolution of the concurrences
and the corresponding probabilities for the two states $\vert\phi_{1}(t)\rangle$ and $\vert\phi_{2}(t)\rangle$.
Here we use the scaled time $\tau=gt$, and for satisfying the condition $\omega_{M}\gg g$, we choose $\omega_{M}=15g$.
We can see from Fig.~\ref{concuandproba} that the concurrences exhibit some oscillations with a sine-cosine envelope,
and that the probabilities show inverse behavior with respect to the concurrence.
When the concurrence is large, the probability is small, and vice versa. So there is a competition between achieving either a large entanglement or a large probability.
At certain times, the concurrence
could be very close to $1$. These times can be determined based on Eq.~(\ref{concurrence}). For the state $\vert\phi_{1}(t)\rangle$,
we choose the time $t$ such that $\cos(\omega_{M}t)=-1$, then $C_{1}(t)=\frac{\sin^{2}(gt)}{4\sin^{2}(gt/2)}$.
If the time $t$ also satisfies the condition $\sin^{2}(gt)\ll1$, then
we have the approximate relations $\sin^{2}(gt)\approx (gt)^{2}$ and $\sin^{2}(gt/2)\approx (gt)^{2}/4$, and $C_{1}(t)\approx1$.
However, the probability for this maximum entanglement state is very small because $P_{1}=\sin^{2}(gt/2)\ll1$.
Similarly, we can determine the times for obtaining a large entanglement state $\vert\phi_{2}(t)\rangle$.
As a compromise, we can choose the times $\tau/\pi= n+1/2$ ($n=0,1,2,...$).
At these times, the concurrences of the states $\vert\phi_{1}(t)\rangle$ and $\vert\phi_{2}(t)\rangle$
are $1/2$, and the two probabilities are also $1/2$.
\begin{figure}[tbp]
\center
\includegraphics[bb=42 136 361 273, width=3.3 in]{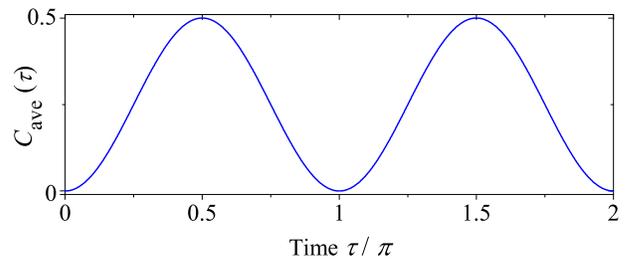}
\caption{(Color online) Dynamics of the average concurrence $C_{\textrm{ave}}(t)$ given in Eq.~(\ref{aveconcu}). The parameters are the same as those in Fig.~\ref{concuandproba}.}
\label{aveconcurrence}
\end{figure}

Based on the above analyses, we now introduce an average concurrence to evaluate the entanglement efficiency:
\begin{eqnarray}
C_{\textrm{ave}}(t)=P_{1}(t)C_{1}(t)+P_{2}(t)C_{2}(t)=\frac{1}{2}\sin^{2}(gt).\label{aveconcu}
\end{eqnarray}
In Fig.~\ref{aveconcurrence}, we illustrate the dynamics of the average concurrence. We can see that the maximum average entanglement
can be achieved at times $\tau=(n+1/2)\pi$ for $n=0,1,2...$. So the shortest time is $\tau_{\textrm{min}}=\pi/2$. In this case, $\beta(\tau_{\textrm{min}})=-2g/\omega_{M}=-2/15$, and the states become
\begin{eqnarray}
\vert\phi_{1}(\tau_{\textrm{min}})\rangle&=&\vert\phi_{2}(\tau_{\textrm{min}})\rangle=\frac{1}{\sqrt{2}}\left[-i\vert 0\rangle_{b_{1}}\vert 0\rangle_{b_{2}}\right.\nonumber\\
&&\left.+\frac{i}{\sqrt{2}}(\vert 0\rangle_{b_{1}}\vert 1\rangle_{b_{2}}-\vert 1\rangle_{b_{1}}\vert 0\rangle_{b_{2}})
\right].\label{genstatephispet}
\end{eqnarray}
This state is a typical two-mode three-component state. Its concurrence is $1/2$, which is in the same order as that for the Bell states.
Since $\vert\phi_{1}(\tau_{\textrm{min}})\rangle=\vert\phi_{2}(\tau_{\textrm{min}})\rangle$, the two mechanical resonator and the two cavity fields are decoupled at these times.

We now give some remarks on the possible realization of this proposal.

(i) In the present method, we consider an ideal case
in which the photon dissipation is neglected. So the time needed for the state generation should be much shorter than the cavity photon lifetime.
If we denote the cavity-field decay rate as $\gamma_{c}$, then the lifetime of a single photon in the
cavity should be $\sim1/\gamma_{c}$. Therefore, the threshold condition for entanglement generation time
is $t=\pi/(2g)\ll1/\gamma_{c}$, which leads to the system parameter condition $g\gg\gamma_{c}$ (the mechanical decay rate $\gamma_{m}$ is much smaller than $\gamma_{c}$).
This is the single-photon strong-coupling condition in optomechanics.
In addition, since we used the condition $\omega_{M}\gg g_{0}$, the system should also work in the deep-resolved-sideband regime $\omega_{M}\gg \gamma_{c}$.

(ii) To prepare the initial state $\vert 1\rangle_{a_{1}}\vert0\rangle_{a_{2}}\vert0\rangle_{b_{1}}\vert0\rangle_{b_{2}}$, the two mechanical mirrors should
be cooled to their ground states, which can be realized by the ground-state cooling method~\cite{Teufel2011,Chan2011}.
For the cavity-field state preparation and measurement, it would be necessary to use the
techniques developed for cavity-QED~\cite{Raimond2001}. The single cavity photon could be loaded by a two-level atom.
An excited atom is assumed to fly through the cavity. By carefully controlling the velocity of the atom, such that a $\pi/2$-rotation transition occurs,
a single photon could be emitted into the cavity.

(iii) The measurement of the cavity-field states can also be realized using atoms. We assume that two atoms in their ground states are flying through the two cavities, respectively.
If one atom is excited, then the corresponding cavity is in a single-photon state, and the other cavity is in its ground state.
Since the single photon has been absorbed by the atom in the measurement, the two cavities after the measurement are in their ground states.
Therefore, in their subsequent dynamics, both mechanical and optical modes only experience free evolution,
so the generated entanglement is preserved. We should point out that, in the single-photon preparation and measurement processes,
the photon-atom coupling strength should be much larger than other coupling scales, so that other physical processes are negligible during the $\pi/2$-rotation.

In summary, we have analytically studied the entanglement dynamics between two mechanical mirrors in a two-cavity optomechanical system.
This entanglement is induced by the radiation pressure of a single photon, which is hopping between the two cavities.
The explicit form of these states was obtained by analytically solving the dynamical evolution of the system. Up to local operations,
these states are two-mode three-component states, and hence we can quantify the entanglement by the concurrence.
It was found that a considerable entanglement can be created when the system works in the single-photon strong-coupling regime and the deep-resolved-sideband regime.
Here, considerable entanglement means that the amount of entanglement of these states is comparable to that of the Bell states.

J.Q.L. would like to thank Zhi-Jiao Deng and Xin-You L\"{u} for discussions. J.Q.L. is supported by the Japan Society for the Promotion of Science
(JSPS) Foreign Postdoctoral Fellowship No. P12503. Q.Q.W. is supported by the Scientific Research Fund of the Education
Department of Hunan Province No.12C0716. F.N. is partially supported by the RIKEN iTHES Project, MURI Center for Dynamic Magneto-Optics, JSPS-RFBR Contract No. 12-02-92100, Grant-in-Aid for Scientific Research (S),
MEXT Kakenhi on Quantum Cybernetics, and the JSPS via its FIRST program.

\end{document}